\documentstyle[11pt,aaspp,natbib,psfig]{article}

\psfigurepath{figs}
\psfull  
\pssilent  

\newcommand{\msun}{M$_{\odot}$}

\def\h2o{\mbox{$\rm H_{2}O$}}
\def\co2{\mbox{$\rm CO_{2}$}}
\def\1a{\mbox{$1/a$}}
\def\las{\mathrel{\hbox{\rlap{\hbox{\lower3pt\hbox{$\sim$}}}\hbox{\raise2pt\hbox{$<$}}}}}
\def\gas{\mathrel{\hbox{\rlap{\hbox{\lower3pt\hbox{$\sim$}}}\hbox{\raise2pt\hbox{$>$}}}}}

\newcounter{saveeqn}

\begin{document}

\title{The stability of planets in the Alpha Centauri system}
\author{Paul A. Wiegert$^1$, Matt Holman$^2$}
\affil{$^1$ Department of Astronomy, University of Toronto, Toronto, Canada\\$^2$Canadian Institute for Theoretical Astrophysics, Toronto, Canada}

\begin{abstract}

This paper investigates the long-term orbital stability of small
bodies near the central binary of the Alpha Centauri system.  Test
particles on circular orbits are integrated in the field of this
binary for 32~000 binary periods or approximately 2.5~Myr. In the
region exterior to the binary, particles with semi-major axes less
than roughly three times the binary's semi-major axis $a_b$ are
unstable. Inside the binary, particles are unstable if further than
0.2~$a_b$ from the primary, with stablility closer in a strong
function of inclination: orbits inclined near $90\deg$ are unstable in
as close as 0.01~$a_b$ from either star.
\end{abstract}

\section{Introduction}

Though the formation of multiple star systems is possibly quite
different from that of single stars like our Sun, it is plausible that
multiple stars also host planetary systems. If this is the case, then
the high frequency of binary and multiple systems implies that such
planetary systems have been created in large numbers in our Galaxy.
However, the question of whether planets might persist for long
periods within such a system remains unanswered.  Alpha Centauri, a
triple system with two of the stars forming a close binary (semi-major
axis 23~AU) and a third orbiting this pair at a much greater distance
(12~000~AU), is extraordinary only in its proximity to the Sun
(1.3~pc). For this reason, it is a prime place to prospect for
planets, and a logical starting point for our theoretical
investigations of the stability of planetary orbits in multiple
systems.

Stability considerations can constrain the locations where planets are
likely to exist. As direct imaging and astrometric techniques are most
suited to detecting planets on large orbits, while spectroscopic
methods are better at detecting small orbits, an understanding of
long-term stability in binary systems can increase the efficiency of
searches for extra-solar planets. We seek regions of phase space where
test particles (planets) could remain for times on the order of the
ages of the stars. More precisely, we will determine those regions in
which planets cannot be stable on such time scales. Our integrations
follow test particles for only a few million years, and thus cannot
assure stability over the $\alpha$~Cen system's probable 5 billion
year age \cite[]{noegremag91}. However, even such relatively short
integrations are sufficient to identify large regions in which single
planets are unstable, and thus cannot exist today.

We adopt a simple, empirical, observationally motivated criterion for
stability. The term ``stable'' will be applied to test particles whose
time-averaged semi-major axis does not vary from its initial value by
more than 5\% over the whole integration, the remainder being termed
``unstable''. Thus our definition of stability excludes planets which
remain bound to the binary, but migrate to larger or smaller orbits,
encompassing only such planets as remain near their initial orbits. We
also compute Lyapunov exponents, which measure the rate of exponential
divergence of nearby orbits, and are correlated with stability
lifetimes.

\section{Method and Models}

The numerical integrations in this paper used the symplectic mapping
for the N-body problem described by \cite{wishol91}.  This technique
is typically an order of magnitude faster than conventional
integration methods and has the additional advantage of showing no
spurious dissipation other than that introduced by roundoff error.
Lyapunov times were computed by evolving a tangent vector associated
with each test particle during the orbit calculations
\cite[]{mikinn94}.  This procedure has two advantages over the common
approach of measuring the Lyapunov time by evolving two nearby
trajectories.  First, using the tangent vector avoids the saturation
and renormalisation problems that accompany the two-trajectory
technique.  Second, the variational method is faster because the most
expensive calculations required (the distances between the test
particle and planets) do not need to be computed twice.

We approach the problem with a simple model which captures the overall
dynamics.  We ignore the distant third star, $\alpha$~Cen~C (Proxima),
as it appears likely that it is not bound to the central binary
\cite[]{anoorlpav94}, and because the perturbations it could inflict
were it bound are extremely small.  The orbit of the central pair
is thus taken to be a fixed Kepler ellipse. The semimajor axis of the
central binary $a_b$ is 23.4~AU, and its eccentricity is 0.52 and the
inclination of its orbit to the plane of the sky is $79\deg$
\cite[]{hei78}. The primary, $\alpha$~Cen~A, has a mass of
1.1~{\msun}; the secondary, $\alpha$~Cen~B, has a mass of 0.91~{\msun}
\cite[]{kamwes78}. Their physical properties are outlined in
Table~\ref{ta:lumetc}.  In the field of this binary we integrate a
battery of massless test particles representing low-mass planets.  As
these particles do not interact with one another, this paper does not
address the stability of multiple planet systems.

\begin{table}
\centerline{
\begin{tabular}{cccccc}  
Star & Mass (M$_{\odot}$) & MK class & $V$ & $M_V$ & L (L$_{\odot}$) \\ \hline 
$\alpha$ Cen A & 1.1 & G2V & --0.01 & 4.37 & 1.6 \\
$\alpha$ Cen B & 0.91& K1V & {~1.33}& 5.71 & 0.45 \\  
\end{tabular}}
\caption{Physical characteristics of $\alpha$~Cen A and B, including their mass, spectral type, apparent and absolute visual magnitude and luminosity \protect\cite[]{kamwes78,bsc,lan92}.}
\label{ta:lumetc}
\end{table}

\section{Initial Conditions}

Test particles are initially placed in circular orbits in two separate
regions. The interior region is centred on the primary, and extends
from 0.01~to~0.5 times the binary semi-major axis $a_b$
(0.23~to~11.7~AU). The exterior region is centred on the barycentre,
and spans $1.5a_b$ to $5a_b$ (35~to~117~AU). Note that the mass
fraction in the secondary (0.45) exceeds the maximum value ($\sim
0.005$ for a binary eccentricity of 0.52, Danby 1964) \nocite{dan64}
at which the $L_4$ and $L_5$ Lagrange points are stable, so no
particles are expected to survive there.

No separate study of the dynamics of orbits centred on $\alpha$~Cen~B
was performed due to the similarity of the masses of the primary and
secondary.  Such a study is expected to produce results qualitatively
very similar to those obtained for orbits centred on $\alpha$ Cen A.

As the central binary has minimum and maximum separations of 0.48$a_b$
and $1.52 a_b$, particles on circular orbits with semi-major axes in
this range suffer close encounters with the secondary, and are
unlikely to be stable.  Thus, no test particles are started in the
semi-major axis range $0.5a_b$ to $1.5a_b$.

The integration is started with the perturber at apastron, and on the
opposite side of the primary from the particles. The particles are
initially in the plane of the binary, but have a range of
inclinations.

Thirteen different inclination values were examined, ranging from
0$\deg$ to 180$\deg$ in 15$\deg$ increments. All particles were
started on circular orbits, relative to the primary in the inner
shell, and relative to the barycentre in the outer one. Particles were
distributed evenly in initial semi-major axis $a$, 50 particles for
each value of the inclination in the inner region ($0.01 a_b$ particle
separation), and 36 particles for each value of the inclination ($0.1
a_b$ particle separation) in the outer one, for a total of 1118
particles in both regions. The integration proceeded for 32~000 binary
periods, approximately 2.5 Myr of simulated time. The time step used
was $3 \times 10^{-3}$ of the binary period in the outer region; in
the inner region, the step size was $10^{-4}$ from $0.11$ to $0.5
a_b$, and $3 \times 10^{-5}$ for the particles with semi-major axes
less than and including 0.1~$a_b$. These step sizes translate into 33
steps per particle orbit at 0.01~$a_b$, 360 per orbit at 0.11~$a_b$
and $610$ per orbit at 1.5~$a_b$.

When integrating particles in the inner region, the mass of the
secondary was grown adiabatically over 500 binary periods in order to
eliminate transients in the particles' motions which would not be
present in a mature planetary system. The time span of 500 binary
periods (roughly 40~000~years) is comparable to the precession period
of test particle orbits with semi-major axes as small as $0.05~a_b$.
The adiabatic growth procedure was found to have little effect on the
final results, and was omitted in the calculations of the exterior
region. In the inner region, 498 of the 650 particles became unstable
during the adiabatic growth phase, with those with the largest orbits
typically being lost first.

\section{Simulations}

\begin{figure}[p]
\vspace*{-0.5in}
\centerline{
\psfig{figure=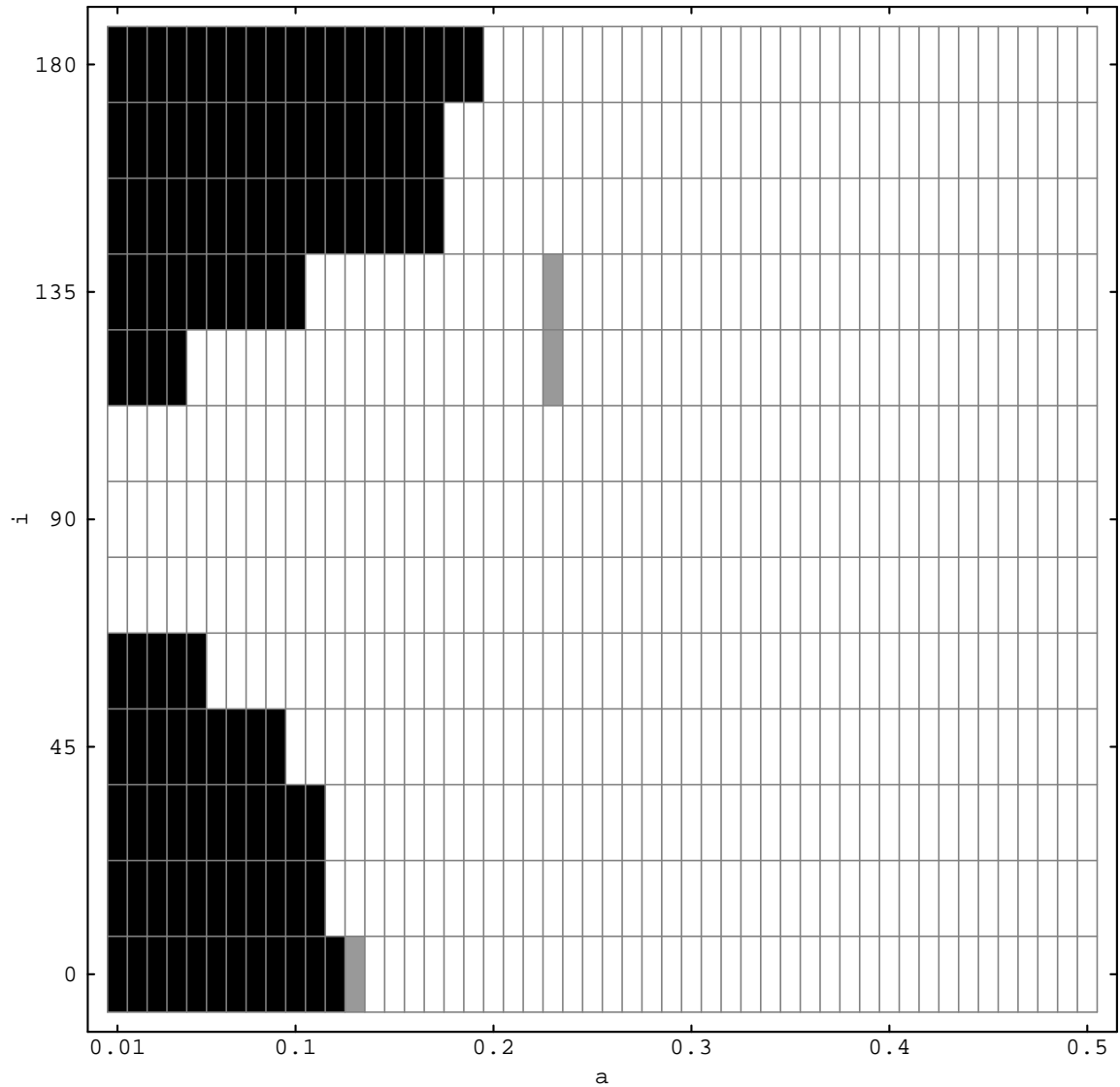,width=6in}}
\vspace*{-1in}
\caption{The change in semi-major axis of test particles in the inner
region of the $\alpha$ Cen binary, on a grid of inclination $i$ and
semi-major axis $a$. A white cell indicates a particle that was
ejected or had a close encounter with the secondary.  Particles which
survived the whole integration time, but whose average semi-major axis
differs from its initial value by more than $5\%$ are indicated by a
grey cell, while a deviation of less than 5\% is indicated by a black
cell.}
\label{fi:innera}
\end{figure}

The inner region proves to be largely unstable over the integration
time scale (Figure~\ref{fi:innera}). Each cell of
Figure~\ref{fi:innera} represents one of the test particles' initial
conditions.  A white cell indicates a particle that was ejected or had
a close encounter (defined to be a passage within 0.25 $a_b$) with the
secondary.  Two other colours indicate those particles that survived
for the entire simulation: those whose time-averaged semi-major axis
deviated from its initial value by less than $5\%$ are indicated in
black, those which deviated by more than $5\%$ are shown in grey.

\begin{figure}[p]
\vspace*{-0.5in}
\centerline{
\psfig{figure=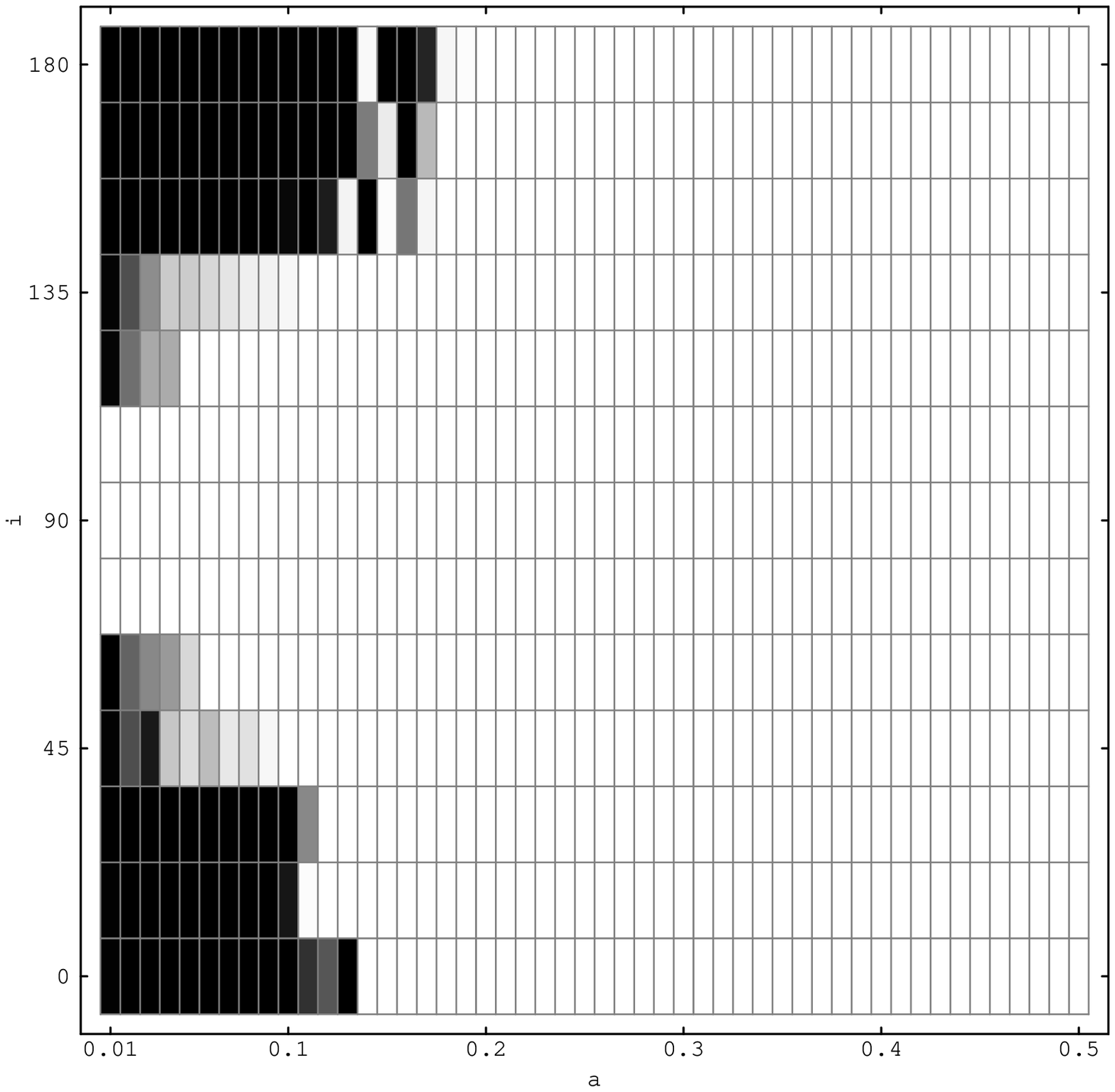,width=6in}}
\vspace*{-1in}
\caption{The Lyapunov time of test particles in the inner region of
the $\alpha$ Cen binary, on a grid of inclination $i$ and semi-major
axis $a$. Black indicates the longest detectable Lyapunov times
(approximately 1200 binary periods), shading to white, the lowest, at
or near zero. Particles which were ejected or which moved by more than
$50\%$ of their initial semi-major axis are grouped with the lowest
Lyapunov times (white).}
\label{fi:innerb}
\end{figure}

The region where particles are longest-lived is close to the primary
as might be expected, but with a wide gap at inclinations between
$60\deg$ and $120\deg$. This gap presumably closes at smaller
distances from the central star. Orbits in the plane of the binary are
stable out to larger distances than those with significant
inclinations. Retrograde orbits survive out to larger radii than
prograde ones, as might be expected from the shorter encounter times
suffered by retrograde orbits and from studies of distant outer planet
satellites \cite[]{hen70}.

There are only three grey particles in Figure~\ref{fi:innera}. The two
grey ones at $a = 0.23 a_b$ survive but migrate out onto larger
orbits outside the binary ($a \sim 5 a_b$). The grey one at
$i=0\deg, a=0.13 a_b$ moves inward slightly to roughly $0.1
a_b$. Overall, the $a$-$i$ plane is divided fairly cleanly into two
parts, one stable and one unstable on million year time scales.

A plot showing Lyapunov times for the inner region appears in
Figure~\ref{fi:innerb}. The plot shows the same general
stable/unstable division as Figure~\ref{fi:innera}. There are,
however, regions in Figure~\ref{fi:innerb} that show Lyapunov times
below the maximum detectable level (which is about 1200 binary orbital
periods, or 0.1~Myr in the inner region), but which have not moved
significantly from their initial positions. As the time scale for
large qualitative changes in test orbits has been observed to be much
longer than the Lyapunov times in some cases \cite[]{lecframur92},
such particles may move away from their initial positions over longer
time scales.

\begin{figure}[p]
\vspace*{-0.5in}
\centerline{
\psfig{figure=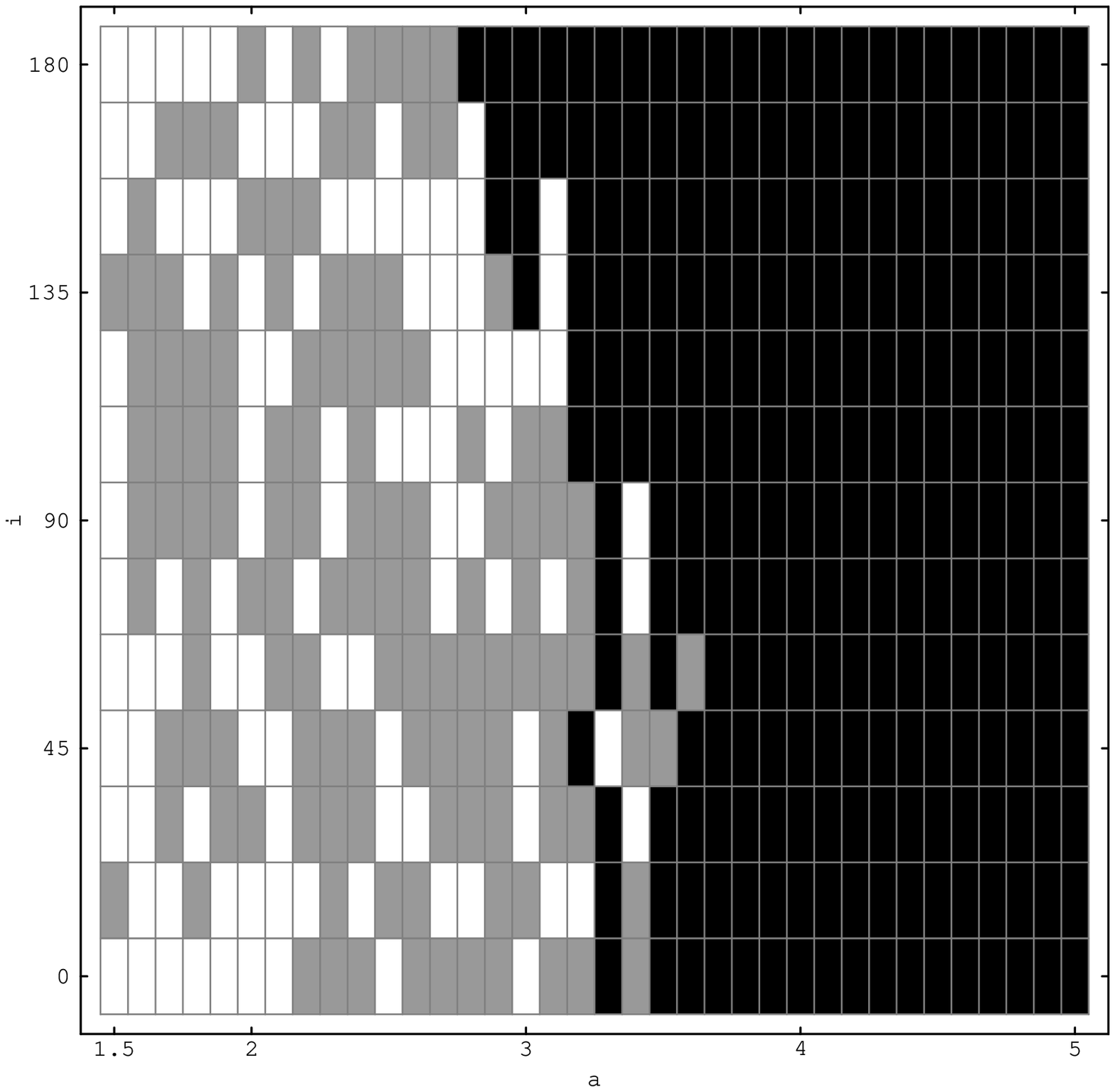,width=6in}}
\vspace*{-1in}
\caption{The change in semi-major axis of test particles in the outer
region of the $\alpha$ Cen binary, on a grid of inclination $i$ and
semi-major axis $a$. The shadings are the same as in
Figure~\protect{\ref{fi:innera}}.}
\label{fi:outera}
\end{figure}

\begin{figure}[p]
\vspace*{-0.5in}
\centerline{
\psfig{figure=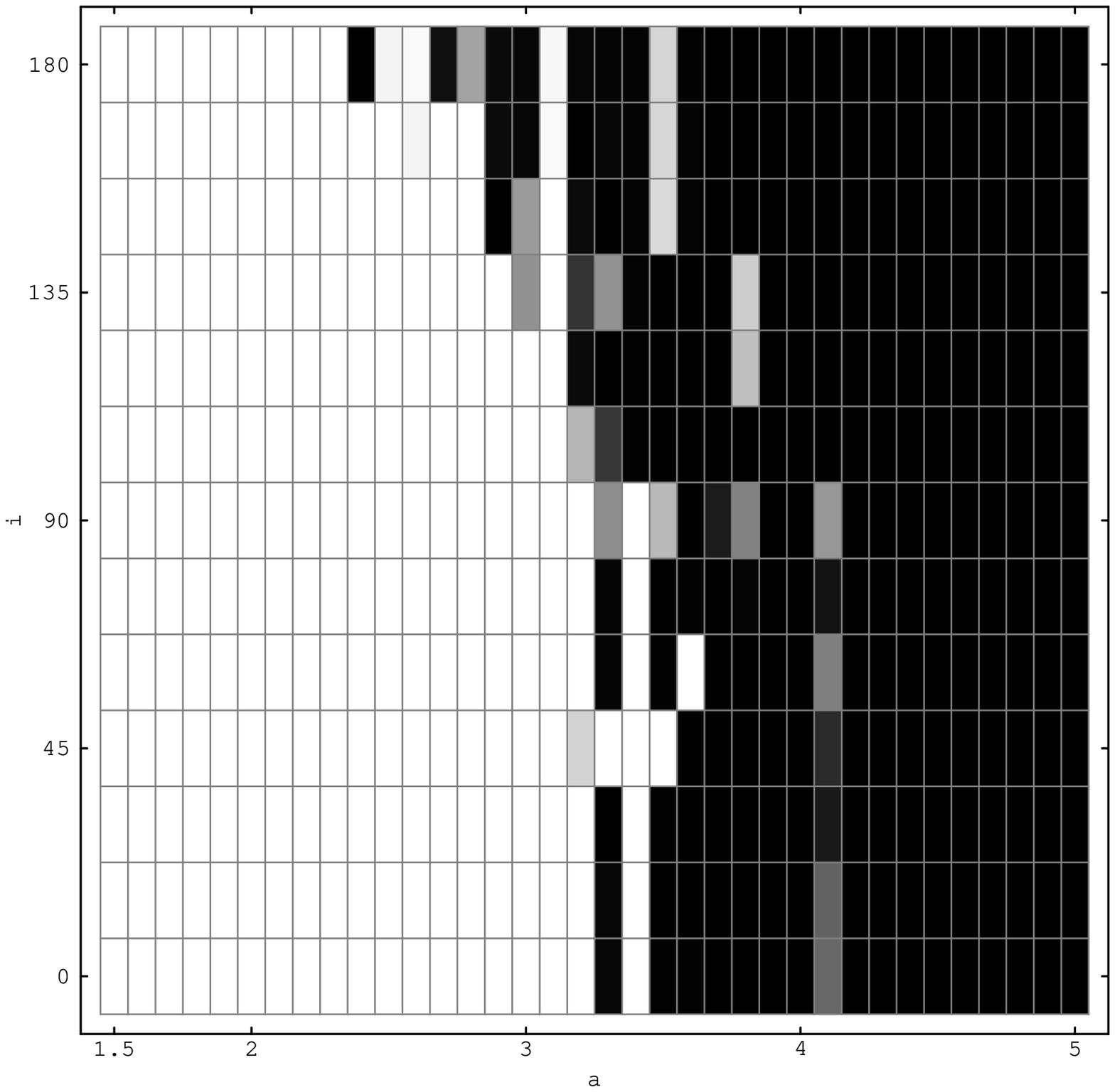,width=6in}}
\vspace*{-1in}
\caption{The Lyapunov time of test particles in the outer region of
the $\alpha$ Cen binary. The shadings are the same as in
Figure~\protect{\ref{fi:innerb}}, except that the maximum Lyapunov
time is now roughly 2800 binary periods.}
\label{fi:outerb}
\end{figure}

Plots analogous to those in Figures~\ref{fi:innera} and
\ref{fi:innerb} but for the outer region are shown in
Figures~\ref{fi:outera} and \ref{fi:outerb}. In
Figure~\ref{fi:outera}, we again see a division of the region into
stable and unstable regions.  Particles within roughly $3 a_b$ are for
the most part unstable, with the stable region reaching further
inwards for retrograde orbits, while orbits outside $4 a_b$ survive
for the length of the integration.

There are many grey cells in Figure~\ref{fi:outera}, indicating
particles which have survived the integration but whose semi-major
axes wander significantly from their initial values. Almost all of
these particles move to larger orbits well outside the binary. Only
one moves to a more tightly bound orbit, and a few along the stability
edge remain within 50\% of their initial semi-major axis, but will
presumably move away on longer time scales.

The behaviour of the Lyapunov times in the outer region, displayed in
Figure~\ref{fi:outerb}, is generally consistent with
Figure~\ref{fi:outera}.  Particles which show only small changes in
semi-major axis generally have longer Lyapunov times, except for a few
isolated particles. It is unclear if these exceptions indicate
isolated pockets of chaotic behaviour associated with narrow
resonances, or possibly chaos on time scales of order or slightly
larger than the maximum detectable Lyapunov time, and which are just
at the edge of detectability in these simulations.  The maximum
detectable Lyapunov time is slightly longer outside than inside, at
roughly $2800$ orbits or 0.2~Myr.

In order to further investigate the stability of planets over even
longer time scales, all particles between 0.05 $a_b$ and 0.15 $a_b$
and at zero inclination were run 50 times longer (1.6 million binary
periods or 130 million years). Particles inside and including that at
0.1 $a_b$ were stable, with no signs of chaos on time scales less than
the maximum detectable Lyapunov time (roughly 50~000 binary periods or
4 million years) while the particles outside 0.1~$a_b$ were all
ejected or suffered close encounters. Comparison of this result with
those in Figures~\ref{fi:innera} and \ref{fi:innerb} indicates that
the inner edge of what the ``stable'' region may be eroded somewhat as
integration times are extended, though the time scale for this effect
and whether or not it will reach arbitrarily far inwards is unclear.

\section{Questions and Comments}

Although these integrations cannot assure the stability of planets on
time scales greater than about a million years, they do identify
important unstable regions. The zone in which planets cannot have
survived since the formation of the $\alpha$ Cen system extends from
at least $0.15 a_b$ (3.5~AU) to $3 a_b$ (70~AU) for orbits which lie
in the binary's orbital plane. Retrograde orbits may be stable as far
as $0.2 a_b$ (4.7~AU) from the primary, and in as far as $2.8 a_b$
(66~AU). Orbits lying perpendicular to the plane are unstable in as
close as $0.01 a_b$ (0.23~AU) from the primary, though smaller stable
orbits are not excluded by our studies.

\begin{figure}[p]
\centerline{
\psfig{figure=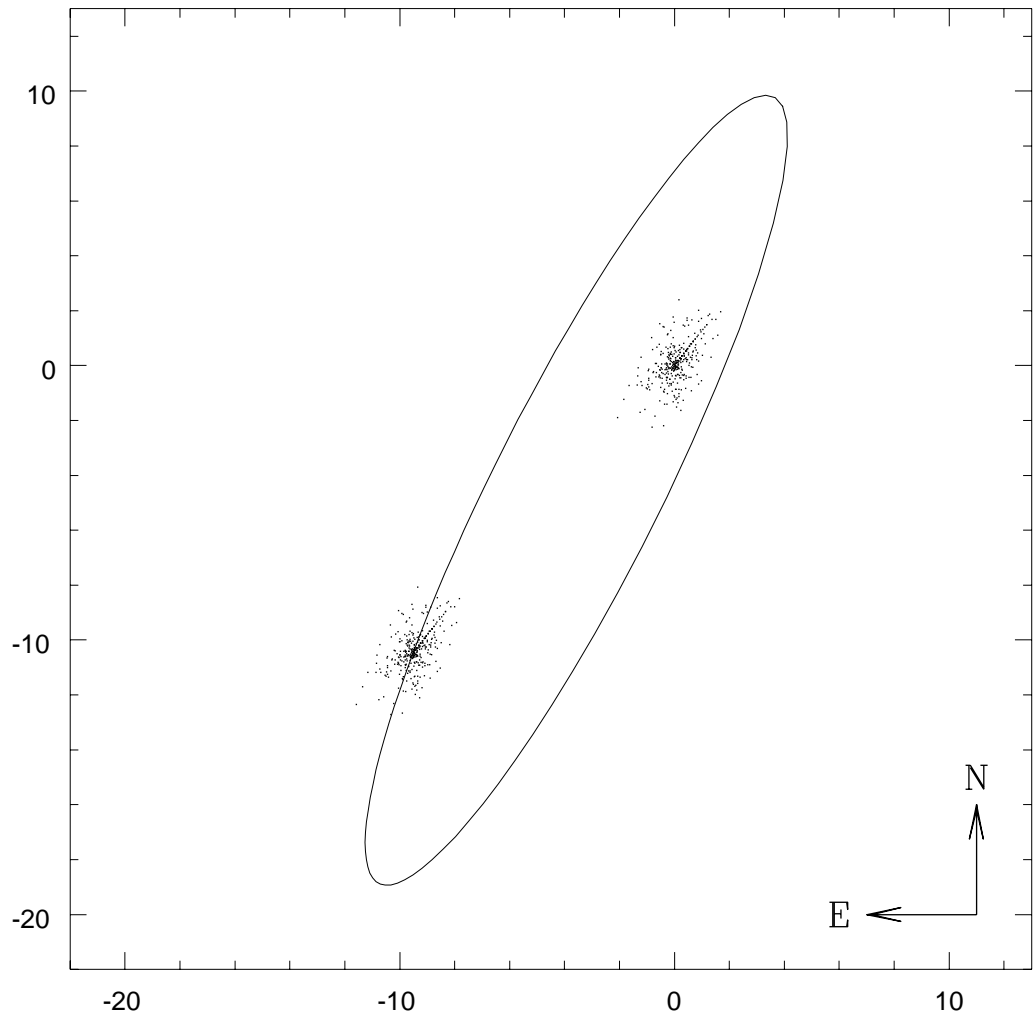,width=6in}}
\caption{The projected density of stable interior planets around the
$\alpha$~Cen binary. The orbit of the secondary is based on
\protect\cite{worhei83}, and the secondary's position is indicated for
the epoch 2000.0. The axes are in arcseconds.}
\label{fi:skyin}
\end{figure}

\begin{figure}[p]
\centerline{
\psfig{figure=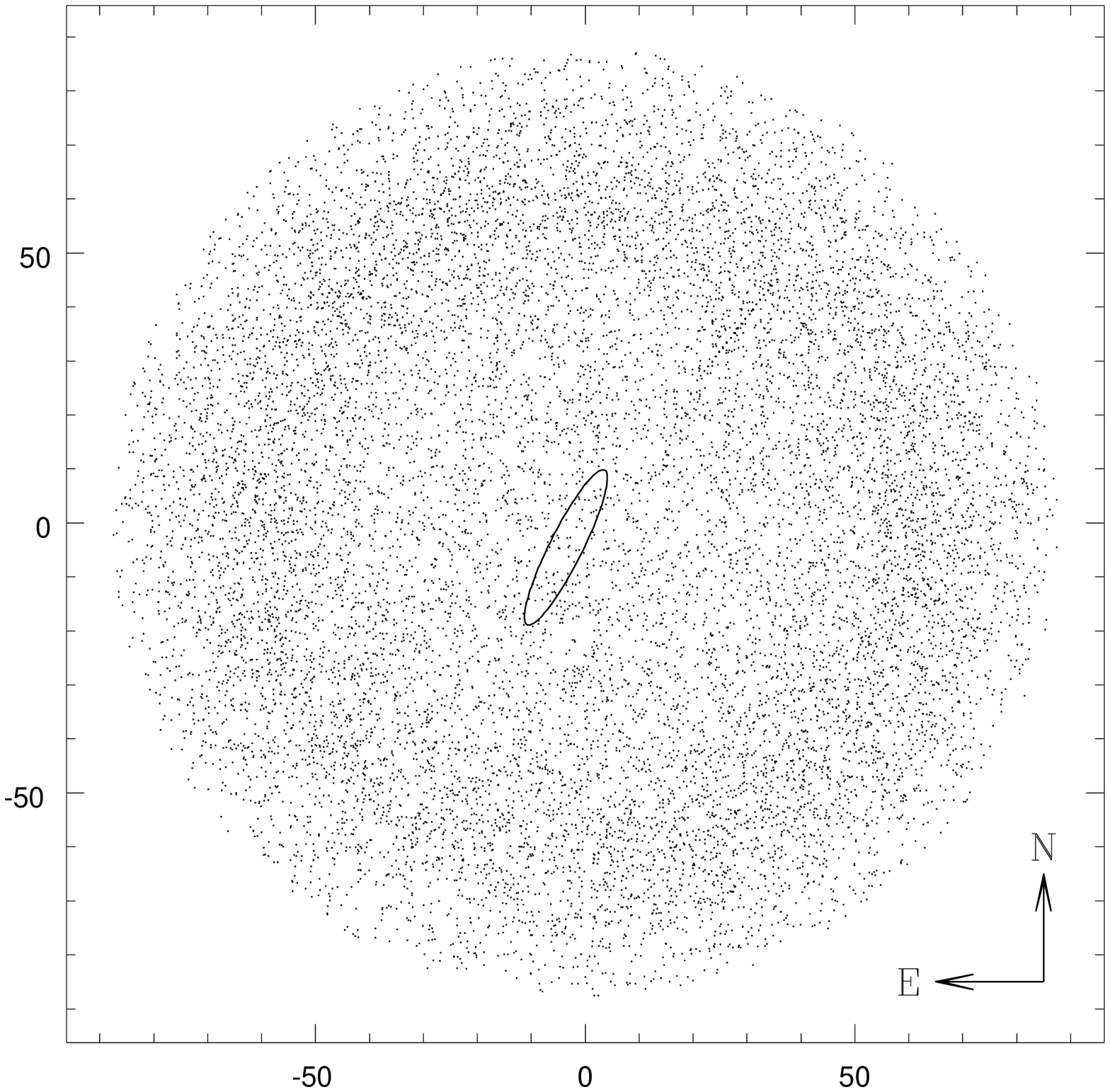,width=6in}}
\caption{The projected density of stable exterior planets around
$\alpha$~Cen with semi-major axes less than 5 $a_b$. The axes are in
arcseconds.}
\label{fi:skyout}
\end{figure}

The stable regions, as seen when projected onto the plane of the sky,
are presented in Figures~\ref{fi:skyin} and \ref{fi:skyout}. The
density of plotted points is proportional to the projected density of
planets in the $\alpha$~Cen system if the phase space corresponding to
the black cells in Figures~\ref{fi:innera} and \ref{fi:outera} is
uniformly populated with circular orbits. 

The habitable zone for planets, as defined by \cite{har79}, lies about
1.2--1.3~AU ($1''$) from $\alpha$~Cen~A. A similar zone may exist
0.73--0.74~AU ($0.6 ''$) from $\alpha$~Cen~B. From our investigations,
it appears that planets in this habitable zone would be stable in the
sense used here, at least for certain inclinations.

Our results are in qualitative agreement with Harrington's (1968; 1972)
\nocite{har68,har72} studies of hierarchical triple star systems.  He
found that the inner binary tended to be unstable when its orbital
plane was perpendicular to the orbital plane of the most distant
member.

\cite{ben88} also investigated the stability of planets in the
$\alpha$~Cen system. He only explored the case of zero planetary
inclination but did allow for non-zero planetary eccentricities.
Benest found that eccentric retrograde orbits were stable over a
greater range of initial distances from the primary than prograde
ones, but found the opposite to be the case for circular orbits. This
result is different from what we observe here, possibly due to the
short duration of his simulations, which lasted only 100 binary
periods, while ours run over three hundred times longer.

The fact that planets would seem to be more stable when in the plane
of the binary's orbit may increase the likelihood of planets existing
in the $\alpha$~Cen system. If one assumes that the planets form and
then remain roughly in the primary's equatorial plane, as they have in
our Solar System, the coincidence of $\alpha$~Cen's equatorial and
orbital planes \cite[]{doylor84,hal94} indicates that, should
planetary formation have proceeded in a manner similar to that in
which it did here, it is plausible that planets might remain in the
system to this day.

\section{Conclusions}

Our studies reveal that much of the region around the central
$\alpha$~Cen binary is unstable. However, there are zones in which
planets on circular orbits could be stable in the $\alpha$~Cen system
on million year time scales. These zones are located both far from ($a
\gas 70$~AU) and near to ($a \las 3$~AU) the primary. Stability is a
strong function of the inclination for interior orbits, less so
exterior orbits.  The inner stable region encompasses Hart's (1979)
habitable zone, however a planet orbiting in the more distant stable
region would be inhospitable to life.

\section{Acknowledgements}

We are indebted to Tsevi Mazeh for helpful discussions, and to Scott
Tremaine for many insightful suggestions and comments on this project.
This research has been funded by the National Science and Engineering
Research Council of Canada.

\bibliographystyle{natbib}
\bibliography{refer_acen}

\end{document}